\documentclass[useAMS]{mn2e}
\usepackage{graphicx}

\def\arcmin{\tt '}

\def\simgt{\ {\raise-.5ex\hbox{$\buildrel>\over\sim$}}\ }

\def\cd{cd$^{-1}$\,}
\def\thop{$\theta$\,Oph}
\def\thopa{$\theta$\,Oph\,A}
\def\thopb{$\theta$\,Oph\,B}
\def\bcep{$\beta$\,Cephei~}

\begin{document}

\title[Asteroseismology of \thop: photometry]
{An asteroseismic study of the \bcep star $\theta$ Ophiuchi: photometric 
results}
\author[G. Handler et al.]
{G. Handler$^{1}$, R. R. Shobbrook$^{2}$, T. Mokgwetsi$^{3}$
\and \\
$^1$ Institut f\"ur Astronomie, Universit\"at Wien, T\"urkenschanzstrasse
17, A-1180 Wien, Austria\\
$^{2}$ Research School of Astronomy and Astrophysics, Australian National 
University, Canberra, ACT, Australia\\
$^{3}$ Theoretical Astrophysics Programme, University of the North-West,
Private Bag X2046, Mmabatho 2735, South Africa}

\date{Accepted 2004 July 17.
  Received 2004 August 13;
  in original form 2004 September 10}
\maketitle
\begin{abstract}

We have carried out a three-site photometric campaign for the $\beta$
Cephei star \thop ~from April to August 2003. 245 hours of differential
photoelectric $uvy$ photometry were obtained during 77 clear nights.
The frequency analysis of our measurements resulted in the detection of
seven pulsation modes within a narrow frequency interval between 7.116 and
7.973 \cd. No combination or harmonic frequencies were found. We performed
a mode identification of the individual pulsations from our colour
photometry that shows the presence of one radial mode, one rotationally
split $\ell=1$ triplet and possibly three components of a rotationally
split $\ell=2$ quintuplet. We discuss the implications of our findings and
point out the similarity of the pulsation spectrum of \thop ~to that of
another \bcep star, V836~Cen.

\end{abstract}


{\section{Introduction}}

Two recent groundbreaking studies have opened up the class of the \bcep
pulsators for asteroseismic investigations. For the star V836 Cen, Aerts
et al. (2003, 2004a) acquired and analysed 21 years of time-resolved
Geneva photometry. They identified the six detected pulsation modes with
their pulsational quantum numbers (the radial fundamental mode, an
$\ell=1$ triplet and two components of a rotationally split $\ell=2$
mode). Consequent seismic modelling (Dupret et al. 2004) allowed the
derivation of constraints on the star's position in the HR diagram and its
convective core size plus demonstrated that its interior rotation is not
rigid.

A second \bcep star, $\nu$\,Eri, was studied with large photometric and
spectroscopic multisite campaigns (Handler et al. 2004, Aerts et al.  
2004b), yielding a total of almost 1200 hours of measurement. The nine
modes detected for this star were identified with the radial fundamental
mode, two $\ell=1$ triplets, one $\ell=1$ singlet and one $\ell=2$ mode
(De Ridder et al. 2004). Seismic modelling (Pamyatnykh, Handler \&
Dziembowski 2004, Ausseloos et al. 2004) demonstrated that the pulsation
spectrum of $\nu$\,Eri cannot be reproduced with standard models, some
convective core overshooting may be required and again non-rigid interior
rotation must be present (with the edge of the convective core rotating
about 3 times faster than the outer layers, consistent with the findings
for V836 Cen). The seismic results indicate that it is possible that the
interior chemical composition of the star is not homogeneous.

After some 15 years of frustration, asteroseismology of opacity-driven
main sequence pulsators has thus finally become reality. The reasons why
some \bcep stars are the first such objects to be studied may be
summarised as follows: their pulsational mode spectra are sufficiently
simple that few possibilities for erroneous or ambiguous mode
identifications occur, yet the observed spectra are fairly complete;
radial modes have been detected for the two abovementioned stars
(substantially reducing the number of possible seismic models); finally,
the applied mode identification methods do work (e.g. see Handler et al.
2003).

The general astrophysical implications of seismic studies of the \bcep
stars are also highly interesting. Since these objects are main sequence
stars between 9 and 17 $M_{\sun}$ (Stankov \& Handler 2005), they are
progenitors of Type II supernovae, which in turn are largely responsible
for the enrichment of the interstellar medium and thus for the chemical
evolution of galaxies. Consequently, if we can trace the evolution of
\bcep stars by sounding their interiors in different evolutionary states,
we are not only able to calibrate stellar structure and evolution
calculations, but could put constraints on the modelling of extragalactic
stellar systems. Therefore it is highly desirable to determine the
interior structure of several \bcep stars.

One of the objects that seems well suited for an asteroseismic study is
\thop. The variability of this bright ($V=3.27$~mag) object has been known
for a long time (Henroteau 1922), and the corresponding period
determinations in the literature are partly controversial. Several authors
(van Hoof \& Blaauw 1958, van Hoof 1962, Briers 1971, Heynderickx 1992)
noted variable shapes of their radial velocity and light curves,
indicating multiperiodicity. However, no consensus on the values of
possible secondary and tertiary periods was reached.

Together with the presence of archival high-resolution spectroscopy (to be
analysed in a companion paper by Briquet et al. 2005), the findings
mentioned above made \thop ~an attractive target for a multisite study.
Consequently, we have carried out a photometric campaign on this star in
mid 2003.

\vspace{4mm}

\section{Observations and reductions}

We acquired single-channel differential photoelectric photometry through
the Str\"omgren $uvy$ filters with three telescopes on three continents
during the months of April to August 2003. The measurements of \thop ~were
obtained with respect to two comparison stars, 44 Oph (HD 157792, A3m,
$V=4.17$) and 51 Oph (HD 158643, A0V, $V=4.81$). Owing to the brightness
of all three objects, some neutral density filters were applied to avoid
damage of the photomultipliers. A short summary of the observations is
given in Table 1. The total time base of our measurements is 124 days.

\begin{table*}
\caption[]{Log of the photometric measurements of \thop. Observatories 
are ordered according to geographical longitude.}
\begin{center}
\begin{tabular}{lccccccc}
\hline
Observatory & Longitude & Latitude & Telescope & \multicolumn{3}{c}{Amount 
of data} & Observer(s)\\
& & & & Nights & h & points & \\
\hline
South African Astronomical Observatory (SAAO) & +20\degr 49\arcmin &
$-$32\degr 22\arcmin & 0.5m & 9 & 34.08 & 135 & TM\\
Fairborn Observatory & $-$110\degr 42\arcmin & +31\degr 23\arcmin & 0.75m 
APT & 44 & 106.68 & 662 & $--$\\
Siding Spring Observatory (SSO) & +149\degr 04\arcmin & $-$31\degr 
16\arcmin & 0.6m & 24 & 104.35 & 506 & RRS\\
\hline
Total & & & & 77 & 245.11 & 1303 \\
\hline
\end{tabular}
\end{center}
\end{table*}

Data reduction was started by correcting for coincidence losses, sky
background and extinction. Nightly extinction coefficients were determined
with the Bouguer method or with the differential technique from the
comparison stars (neither showed any variability during our measurements).
As the comparison stars are considerably cooler than the variable,
second-order colour extinction coefficients were also determined. We found
colour extinction corrections to be necessary for the $u$ data from SSO
and the APT and applied them correspondingly.

We then determined the mean $u, v, y$ zeropoints between the comparison
star magnitudes and used them to combine the measurements of 44 and 51 Oph
to a curve that was assumed to reflect the effects of transparency and
detector sensitivity changes only. Consequently, these combined time
series were binned into intervals that would allow good compensation for
the above mentioned nonintrinsic variations in the target star time series
and were subtracted from the measurements of \thop. The binning minimises
the introduction of noise in the differential light curve of the target.

The timings for this differential light curve were heliocentrically
corrected as the next step. Finally, the photometric zeropoints of the
different sites, which may not be quite the same because of slightly
different wavelength responses of the individual instrumental systems
combined with the different colours of the variable and the comparison
stars, were set to zero. The resulting final combined time series was
subjected to frequency analysis; we show some of our light curves of \thop
~in Fig.\,1. We note that the amplitude of the light variation is 
modulated, but its shape is not; it is always sinusoidal.

The accuracy of the differential light curves of the comparison stars was
4.8 mmag in the $u$ filter, 4.3 mmag in $v$ and 3.9 mmag in $y$ per single
data point. These rather high values are mostly caused by the high air
mass of \thop ~and unstable weather conditions during the measurements at
Fairborn Observatory.

\begin{figure}
\includegraphics[width=88mm,viewport=00 00 318 450]{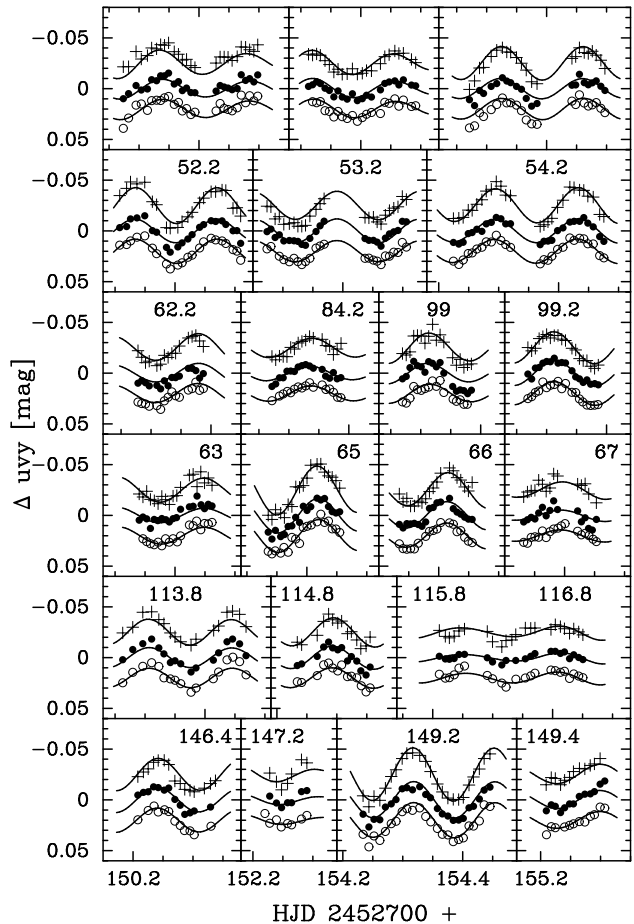}
\caption[]{Some light curves of \thop. Plus signs are data in the 
Str\"omgren $u$ filter, filled circles are our $v$ measurements and open 
circles represent Str\"omgren $y$ data. The full line is a fit composed of 
all the periodicities detected in the light curves (Table 2). The upper 
two panels are measurements from SSO, the middle two are from 
the APT and the lower two from SAAO. The amount of data shown here is 
about one third of the total.}
\end{figure}

\section{Frequency analysis}

Our frequency analyses were performed with the program {\tt PERIOD 98}
(Sperl 1998). This package applies single-frequency power spectrum
analysis and simultaneous multi-frequency sine-wave fitting, and also
includes advanced options.

We started by computing the Fourier spectral window of the final light
curves in each of the filters. It was calculated as the Fourier transform
of a single noise-free sinusoid with a frequency of 7.116 \cd (the
strongest pulsational signal of \thop) and an amplitude of 10 mmag sampled
in the same way as were our measurements. The upper panel of Fig.\,2
contains the result for the $y$ data. Any alias structures that would
potentially mislead us into incorrect frequency determinations are
reasonably low in amplitude due to our multisite coverage.

\begin{figure}
\includegraphics[width=88mm,viewport=00 00 270 565]{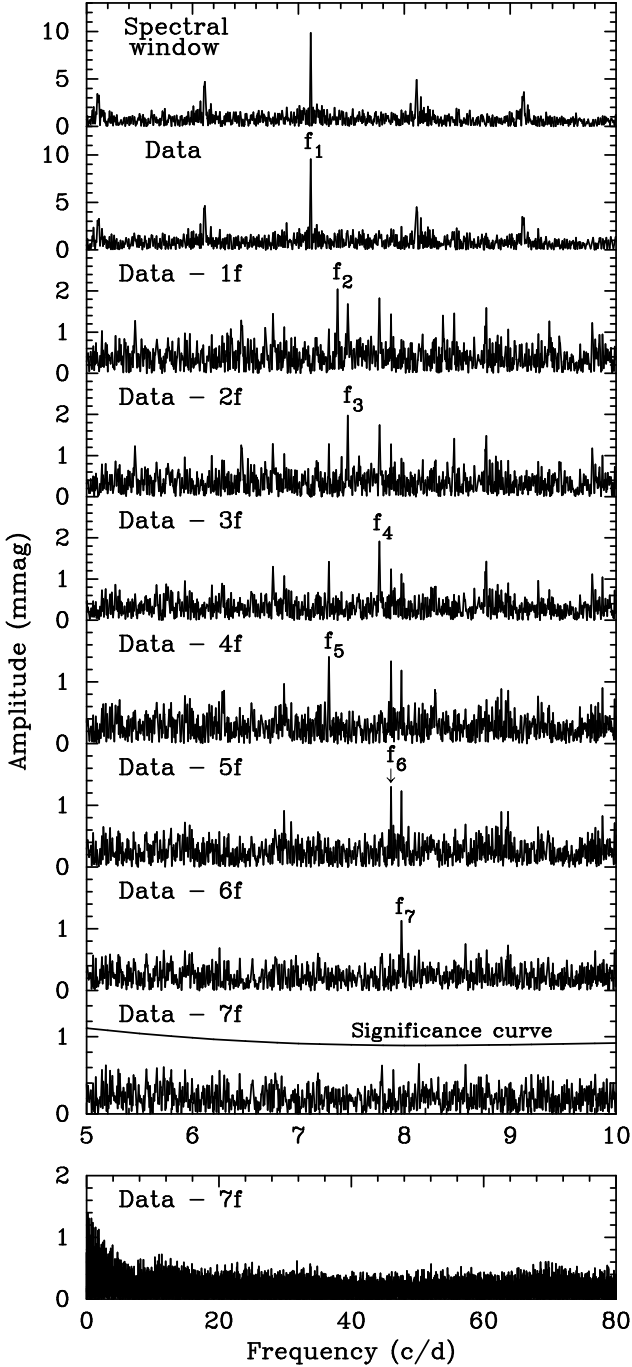}
\caption[]{Amplitude spectra of \thop. The uppermost panel shows 
the spectral window of the data, followed by the periodogram of the data. 
Successive prewhitening steps are shown in the following panels; note 
their different ordinate scales. The second lowest panel contains a 
significance curve; any peak to be regarded as real by us must exceed it. 
The lowest panel shows the residual amplitude spectrum in a wider 
frequency range, containing no evidence for further periodicities in our 
data.} 
\end{figure}

We proceeded by computing the amplitude spectra of the data itself (second
panel of Fig.\,2). The signal designated $f_1$ dominates. We prewhitened
it by subtracting a synthetic sinusoidal light curve with a frequency,
amplitude and phase that yielded the smallest residual variance,
and computed the amplitude spectrum of the residual light curve (third
panel of Fig.\,2).

This resulted in the detection of a second signal ($f_2$). We then
prewhitened a two-frequency fit from the data using the same optimisation
method as before, and continued this procedure (further panels of Fig.\,2)
until no significant peaks were left in the residual amplitude spectrum.

We consider an independent peak statistically significant if it exceeds an
amplitude signal-to-noise ratio of 4 in the periodogram (see Breger et
al.\,1993). The noise level was calculated as the average amplitude in a 5
\cd interval centred on the frequency of interest; the final detection
limit corresponding to $S/N=4$ is shown as a significance curve in the
second lowest panel of Fig.\,2.

We repeated the prewhitening procedure with the $u$ and $v$ data
independently and obtained the same frequencies within the observational
errors. We then determined final values for the detected frequencies by
averaging the values from the individual filters. The pulsational
amplitudes were then recomputed with those frequencies. We regard this
solution as representing our data set best; the result is listed in 
Table\,2.

\begin{table}
\caption[]{Multifrequency solution for our time-resolved photometry of
$\theta$~Oph. The signals are ordered according to their frequencies, but
labelled in the order of detection. Formal error estimates (following
Montgomery \& O'Donoghue 1999) are listed for the individual frequencies;
formal errors on the amplitudes are $\pm$ 0.20 mmag in $u$, $\pm$ 0.18
mmag in $v$ and $\pm$ 0.16 mmag in $y$. The S/N ratio quoted is for the
$y$ filter data.}
\begin{center}
\scriptsize
\begin{tabular}{lccccc}
\hline
ID & Freq. & $u$ Ampl. & $v$ Ampl. & $y$ Ampl. & $S/N$ \\
 & (\cd) & (mmag) & (mmag) & (mmag) & \\
\hline
$f_1$ & 7.11600 $\pm$ 0.00008 & 12.7 & 9.2 & 9.4 & 41.4 \\
$f_5$ & 7.2881 $\pm$ 0.0005 & 2.1 & 1.5 & 1.4 & 6.4\\
$f_2$ & 7.3697 $\pm$ 0.0003 & 3.6 & 2.9 & 2.4 & 10.8\\
$f_3$ & 7.4677 $\pm$ 0.0003 & 4.7 & 2.4 & 2.3 & 10.2\\
$f_4$ & 7.7659 $\pm$ 0.0003 & 3.4 & 2.3 & 2.1 & 9.7 \\
$f_6$ & 7.8742 $\pm$ 0.0005 & 2.3 & 1.8 & 1.3 & 5.8\\
$f_7$ & 7.9734 $\pm$ 0.0005 & 2.4 & 1.6 & 1.2 & 5.6\\
\hline
\end{tabular}
\normalsize
\end{center}
\end{table}

\thop ~has also been observed by the HIPPARCOS satellite (ESA 1997). We
reanalysed the corresponding photometry of the star and find a main
frequency of 7.11605 $\pm$ 0.00002 \cd in these measurements, consistent
with the result from our data within the errors. An analysis of our
measurements combined with those by HIPPARCOS allows us to refine the
value of the dominant frequency to 7.116015 $\pm$ 0.000002 \cd; aliases
are outside the quoted errors for each individual determination. No 
amplitude variations of the strongest mode seem to have occurred between 
the HIPPARCOS measurements and ours; the other signals are not detected in 
the space-based photometry.

The residuals from the multifrequency solution in Table\,2 were searched
for additional candidate signals that may be intrinsic. We have first
investigated the residuals in the individual filters, then analysed the
averaged residuals in the three filters (whereby the $u$ data were divided
by 1.5 to scale them to amplitudes and rms scatter similar to that in the
other two filters), and found no evidence for additional significant
periodicities in any case. The residuals between light curve and fit are
5.2, 4.6 and 4.1 mmag per single $u, v, y$ point, respectively and are
thus somewhat higher than the accuracy of the differential comparison star
data, suggesting that additional, presently undetected, frequencies could
be present.

We can now confront the results of our frequency analysis with those in
the literature. The frequency of the dominant signal is consistent with
all the earlier studies except Henroteau (1922), taking into account some
slight (evolutionary?)  frequency variability with respect to Brier's
(1971) study. Concerning the remaining frequencies, we note that the
resonance period of 0.137255\,d found by van Hoof (1962) is consistent
with our signal $f_5$. On the other hand, none of the secondary or
harmonic frequencies claimed by Heynderickx (1992) can be reconciled with
our data. We suspect this is due to the small amount of data available to
this author. Finally, we note that the $y$ amplitude of our analysis is
consistent with that in Heynderickx' (1992) Walraven $V$ data, i.e. no
amplitude variations seem to have occurred between the years 1987 and
2003.

\section{Mode identification}

Our three-colour photometry gives us the possibility of deriving the
spherical degree $\ell$ of the individual pulsation modes from an analysis
of the colour amplitudes. This involves a comparison of the observed
amplitudes with those predicted by models and first requires knowledge of
the star's position in the HR diagram. 

However, \thop ~is not a single star. Besides its low-mass spectroscopic
companion discovered by Briquet et al.\ (2005), it is also a Speckle
binary (McAlister et al.\ 1993). Shatsky \& Tokovinin (2002) determined a
$K$ magnitude difference of 1.09 mag between the two components and argued
that the companion to the $\beta$ Cephei star (hereinafter called \thopa)  
is physical. From the standard relations by Koornneef (1983) we can infer
that the Speckle companion (hereinafter called \thopb) is 1.33 mag fainter
in $V$ and must thus be a B5 main sequence star.

To determine the effective temperature and luminosity of \thopa, we must
take the contribution of \thopb ~to the total light into account. We use
the standard Str\"omgren photometry by Crawford, Barnes \& Golson (1970),
and adopt the mean $V$ magnitude from the Lausanne Photometric data base
({\tt http://obswww.unige.ch/gcpd/gcpd.html}, $V=3.266$) for the system,
and then reproduce it and the Str\"omgren $c_1$ index, which is a measure
of the stars' effective temperatures, with the help of the standard
relations by Crawford (1978), and after dereddening. The results are shown
in Table~3.

\begin{table}
\caption[]{Johnson V magnitude and Str\"omgren colour indices of 
the \thop ~system.}
\begin{center}
\begin{tabular}{lccccc}
\hline
 & $V$ & $b-y$ & $m_1$ & $c_1$ & $\beta$ \\
\hline
Observed & 3.266 & -0.092 & 0.089 & 0.104 & 2.617 \\
Dereddened & 3.223 & -0.102 & 0.092 & 0.102 & 2.617\\
\thopa & 3.546 & -0.109 & 0.08 & 0.07 & 2.640 \\
\thopb & 4.876 & -0.089 & 0.095 & 0.25 & 2.684\\
Combined & 3.266 & -0.104 & 0.083 & 0.105 & 2.650\\
\hline
\end{tabular}
\end{center}
\end{table}

The observations are reasonably well matched, with the exception of the 
luminosity sensitive $\beta$ parameter. However, this is not a severe 
problem as we will derive the star's luminosity from its parallax. We also 
note that $\beta$ measurements by other authors are closer to our 
calculated results as the results from Crawford et al.\ (1970).

The calibration by Napiwotzki, Sch\"onberner \& Wenske (1993) applied to
the Str\"omgren indices listed in Table~3 then results in $T_{\rm eff} =
22900 \pm 900$~K and $M_v = -2.5 \pm 0.5$ for \thopa, and in $T_{\rm eff}
= 18400 \pm 700$~K and $M_v = -1.3 \pm 0.5$ for \thopb. The analysis of
IUE spectra by Niemczura \& Daszy{\'n}ska-Daszkiewicz (2004) yielded
$T_{\rm eff} = 22200 \pm 850$~K (consistent with the combined contribution
of both system components) $\log g = 3.77$ and $[M/H]=-0.15\pm0.12$.
Finally, the HIPPARCOS parallax of the system (5.79 $\pm$ 0.69 mas)
results in $M_v = -2.6 \pm 0.3$ for \thopa ~and $M_v = -1.3 \pm 0.3$ for
\thopb, respectively, consistent with the result from Str\"omgren
photometry. The tables by Flower (1996) then yield bolometric corrections
of $-$2.2 and $-1.7$~mag, respectively, and thus $M_{\rm bol} =-4.8 \pm
0.5$ for \thopa ~as well as $M_{\rm bol} =-3.0 \pm 0.4$ for \thopb. We
show the positions of the \thop ~components in the HR diagram derived in
this way in Fig.\,3. It becomes clear that the observed pulsations must 
originate from \thopa ~only.

\begin{figure}
\includegraphics[width=99mm,viewport=5 00 305 260]{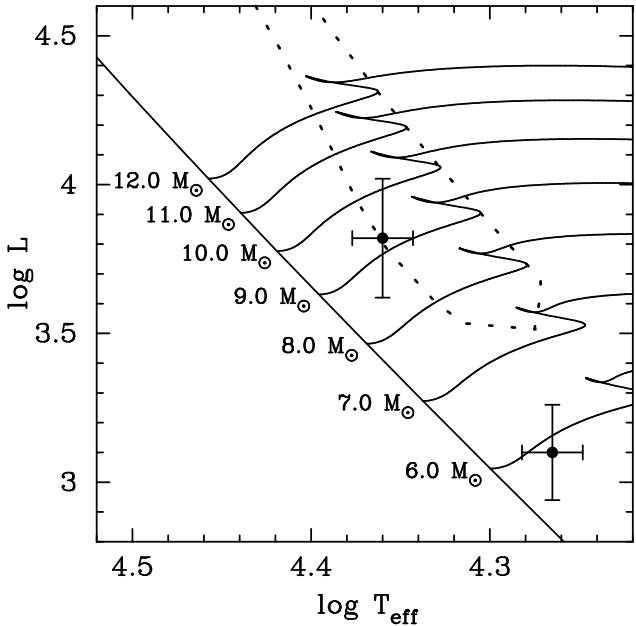}
\caption[]{The position of \thopa ~and \thopb ~in the theoretical HR
diagram. Some stellar evolutionary tracks labelled with their masses (full
lines) and the theoretical borders of the $\beta$ Cephei star instability
strip (Pamyatnykh 1999, dashed lines) are included for comparison. All 
the theoretical results are for a metal abundance of $Z=0.015$. \thopa 
~is located within the instability strip, whereas \thopb ~is not.}
\end{figure}

To derive mode identifications for the pulsations of \thopa, we have
computed theoretical colour amplitudes for modes of $0 \leq \ell \leq 4$
for models with masses between 8.5 and 10 $M_{\sun}$ (in steps of 0.5
$M_{\sun}$), effective temperatures in the range of $4.34 \leq \log T_{\rm
eff} \leq 4.38$ and $Z=0.015$. We first computed stellar evolutionary
models by means of the Warsaw-New Jersey evolution and pulsation code
(described, for instance, by Pamyatnykh et al. 1998). Then we derived the
pulsational amplitudes of such models in the parameter space constrained
above following Balona \& Evers (1999). A range of theoretical frequencies
of 6.5 \cd $\leq f \leq$ 8.5 \cd was examined to allow for some nonradial
mode splitting. Phase shifts between the light curves in the individual
filters were not considered, as no such shifts were observationally found
significant even at the 2$\sigma$ level. We show a comparison of the
observed and theoretical amplitude ratios of three modes in Fig.\,4.

\begin{figure}
\includegraphics[width=85mm,viewport=5 00 255 415]{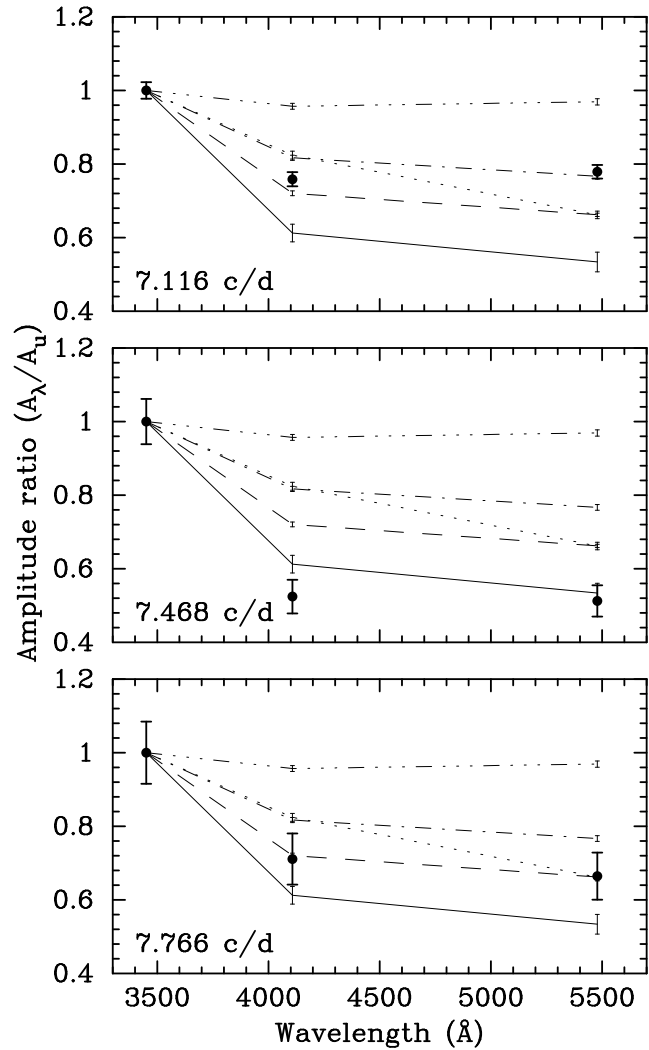}
\caption[]{Observed and theoretical uvy amplitude ratios (lines) for three
modes of \thopa ~and $0\leq\ell\leq4$. Amplitudes are normalised to unity
at u. The filled circles with error bars are the observed amplitude
ratios. The full lines are theoretical predictions for radial modes, the
dashed lines for dipole modes, the dashed-dotted lines for quadrupole
modes, the dotted lines for modes of $\ell=3$ and the
dashed-dot-dot-dotted lines are for $\ell=4$. The small error bars denote
the uncertainties in the theoretical amplitude ratios. The upper panel is
for mode $f_1$, the middle one for $f_3$, and the lower one for $f_4$.}
\end{figure}

We note that we took the contribution of \thopb ~to the total light of the
system into account when determining the observed amplitude ratios; we
found that \thopb ~contributes some 23\% to the total flux in
Str\"omgren~$y$, 22\% in $v$ and 19\% in $u$.

The reliability of the mode identifications in Fig.\ 4 are not easy to
judge. Whereas the modes $f_3$ and $f_4$ can be identified with $\ell=0$
and $\ell=1$, respectively, the situation is less clear for mode $f_1$
(upper panel of Fig.\,4) where the $u/v$ amplitude ratio points towards an
$\ell=1$ mode, but the $u/y$ amplitude ratio suggests $\ell=2$. Similar
problems have been found for other modes that are not shown in this
figure. We believe that the reason for these problems is a combination of
several factors, for instance possible systematic errors in the
determination of some of the pulsational amplitudes (which would be
particularly severe in $u$), the uncertainties of the star's position in
the HR diagram, its poorly constrained surface metallicity (see Dupret
et al. 2004 for a discussion of the latter), and the influence of the 
light of \thopb.

We have therefore chosen an alternative approach that appears more
objective. We calculated the ratio of the individual $u, v, y$ amplitudes
with respect to their mean with the hope of compensating largely for
systematic errors in the amplitude determinations. Then we compared these
ratios to the theoretical ones treated in the same way by means of a
$\chi^2$ analysis, similar to Balona \& Evers (1999) and
Daszy{\'n}ska-Daszkiewicz, Dziembowski \& Pamyatnykh (2003), but
disregarding the pulsational phases since they do, in our case, contain no
information on the type of the modes as argued before. The behaviour of
$\chi^2$ depending on $\ell$ computed this way is shown in Fig.\,5 for all
modes.

\begin{figure}
\includegraphics[width=85mm,viewport=5 05 245 530]{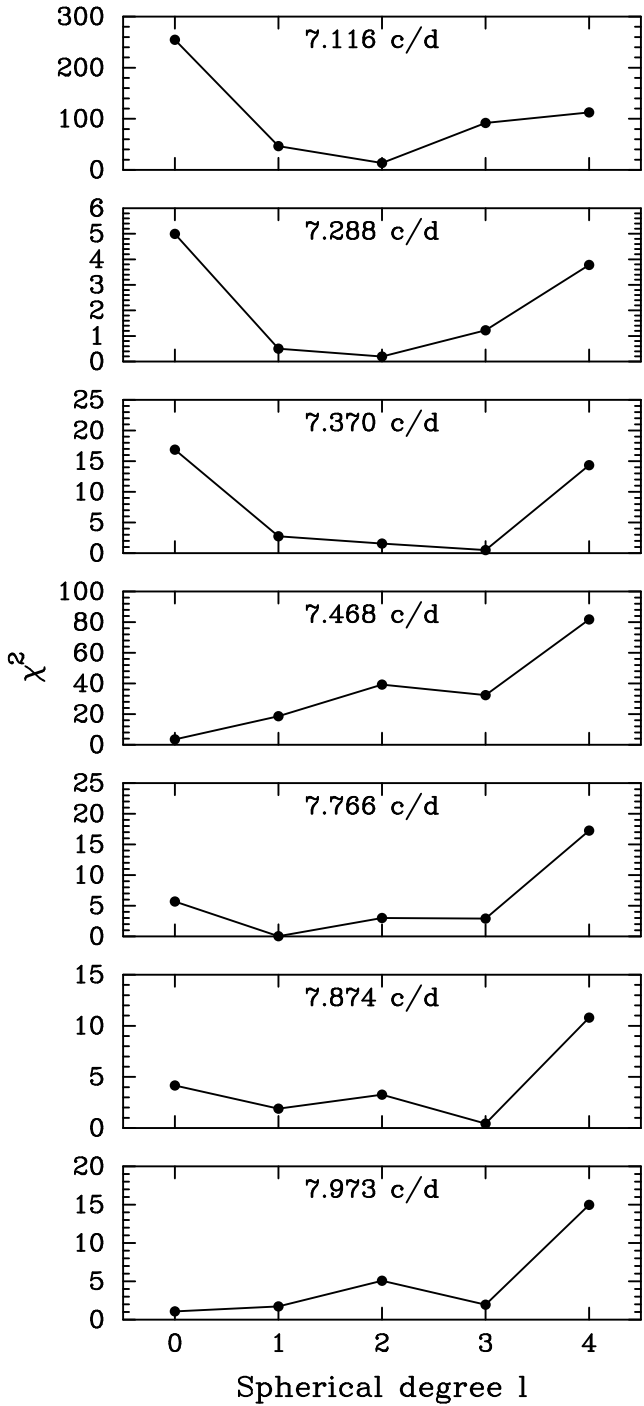}
\caption[]{Mode typing for \thopa ~by means of the $\chi^2$ method.}
\end{figure}

Because of the systematic errors that may affect our mode identification,
we believe that we cannot interpret the results in Fig.\,5 in a strict
statistical sense (i.e. by comparing the observational $\chi^2$ values to
the critical values of a $\chi^2(3)$ distribution and then assign
confidence levels to the derived mode identifications), but that we can
use them to {\it eliminate} some $\ell$ values in the identification
process. The $\ell$ assignments we cannot rule out this way are listed in
Table 4.

\begin{table}
\caption[]{Possible $\ell$ identifications of the individual modes of 
\thopa ~from our $\chi^2$ analysis.}
\begin{center}
\begin{tabular}{lcc}
\hline
ID & Freq. & $\ell$ \\
 & (\cd) & \\
\hline
$f_1$ & 7.11600 & 2 or 1\\
$f_5$ & 7.2881 & 2, 1 or 3\\
$f_2$ & 7.3697 & 3, 2 or 1\\
$f_3$ & 7.4677 & 0 \\
$f_4$ & 7.7659 & 1 \\
$f_6$ & 7.8742 & 3 or 1\\
$f_7$ & 7.9734 & 0, 1 or 3\\
\hline
\end{tabular}
\end{center}
\end{table}

Our mode identifications are not very satisfactory at this point, but
fortunately we can use other clues to constrain them further. Firstly,
since $f_3$ is clearly a radial mode, we can rule out that $f_7$ is also
radial because the frequency ratio of these two modes ($f_3/f_7=0.9366$) 
is considerably larger than any period ratio of low-order radial modes in
a \bcep star can be.

Secondly, Heynderickx, Waelkens \& Smeyers (1994) have identified $f_1$ as
an $\ell=2$ mode from their Walraven photometry. We have checked this
identification with our method (again taking the contribution of \thopb 
~to the total light into account) and also find $\ell=2$ to be clearly the
best match between observed and theoretical colour amplitude ratios.
Thirdly, Daszy{\'n}ska-Daszkiewicz et al. (2002) demonstrated that modes
of odd $\ell$, starting with $\ell=3$, suffer heavy geometric cancellation
in photometric observations of \bcep stars using filters. For instance, an
$\ell=3$ mode of the same intrinsic amplitude as an $\ell=2$ mode will
have only $\sim 1/10$ of its photometric amplitude in the $u, v, y$
filters. We therefore disregard all the possible $\ell=3$ identifications
in Table 4 as well.

Thus we have arrived at unique $\ell$ identifications for five of the
seven modes we detected: $f_1$ is $\ell=2$, $f_3$ is radial, and $f_4,
f_6, f_7$ are all $\ell=1$. $f_2$ and $f_5$ can be either $\ell=1$ or 2. 
As the last step, we examine the frequency spectrum of \thopa
~(schematically plotted in Fig.\,6) for the presence of structures that
may be useful for further constraining the mode identifications.

\begin{figure*}
\includegraphics[width=184mm,viewport=00 05 552 155]{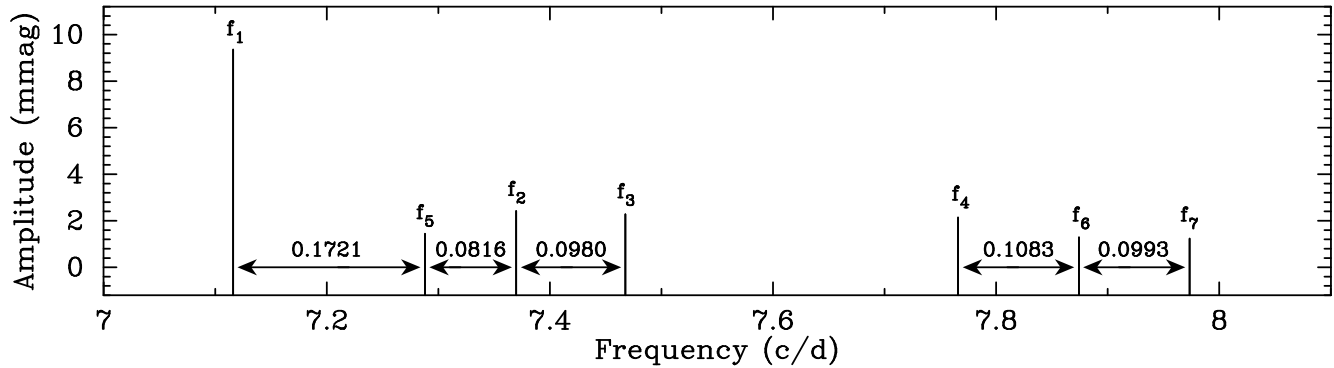}
\caption[]{Schematic amplitude spectrum of \thop. The frequency
differences between some of the modes are indicated.}
\end{figure*}

Indeed, some interesting features can be discerned. The $\ell=1$ modes
$f_4, f_6$ and $f_7$ form a frequency triplet that is almost equally
spaced. However, there is a slight asymmetry, and it is exactly in the 
sense expected for nonradial $m$-mode splitting due to the second-order 
effects of rotation. We therefore believe that $f_4, f_6$ and $f_7$ are 
indeed a rotationally split triplet of $\ell=1$ modes.

The remaining four modes are also grouped together. Intriguingly, the
spacing between $f_2$ and $f_5$ is approximately half the frequency
difference of $f_5$ and the $\ell=2$ mode $f_1$. Again, the asymmetry of
this hypothesised $f_1, f_5, f_2$ multiplet is consistent with the
second-order effects of rotation. $f_3$ does not fit this pattern, but
this is not surprising as we have already identified it as a radial mode.  
The first-order rotational splitting obtained from the $f_1, f_5, f_2$
multiplet is very similar to the splitting of the $f_4, f_6$ and $f_7$
triplet. Thus we suspect that $f_1, f_5$ and $f_2$ are part of a
rotationally split $\ell=2$ quintuplet with two components yet undetected.

Assuming that the mean splitting of the $\ell=1$ triplet is a good
approximation of the surface rotation frequency of \thopa ~(i.e.
neglecting effects of the Coriolis force and possible differential
internal rotation), we derive a rotation period of 9.6 days. The absolute
magnitude and effective temperature of the star, as determined at the
beginning of this section, result in a radius of $5.2\pm1.3$~R$_{\odot}$,
and hence in a surface rotation velocity of $27\pm7$ km/s. As the measured
projected rotational velocity of \thopa ~is about 30 km/s (e.g. Abt,
Levato \& Grosso 2002), there is a chance that we see the star close to
equator-on.

\section{An eclipsing binary?}

If we indeed saw \thopa ~equator-on, it can be suspected that the
spectroscopic companion discovered by Briquet et al.\ (2005) may cause
eclipses. The binary orbit derived by these authors leads to an ephemeris
for the times of primary and secondary minimum, respectively. M. Briquet
(private communication) predicts
\begin{displaymath}
t_I = HJD~ 2451811.002 + i \times 56.712
\end{displaymath}
\begin{displaymath}
t_{II} = HJD~ 2451834.599 + i \times 56.712
\end{displaymath}
where i is the number of orbital revolutions since epoch zero. Given the
total mass of the spectroscopic binary system ($\sim 10$\,M$_{\odot}$),
the orbital period and the radius of the primary determined above, we can
also estimate the maximum duration of a possible eclipse, amounting to
$17\pm4$\,h.


Although we found no obvious evidence for eclipses in our photometric
measurements, we folded our data according to this ephemeris and searched
them again for possible eclipses. As it turns out, we have no measurements
whatsoever during or even near the predicted times of primary minimum,
which is not surprising given that our orbital coverage is only $\sim$ 17
per cent. We do have data around the expected times of secondary eclipse,
but none is found, which is also not a surprise since the secondary of the
spectroscopic binary would much less luminous than \thopa ~if we see the
orbital plane (close to) edge-on, and consequently the depth of a
secondary eclipse would be too small to be detected. We again conclude
that there are no eclipses in our photometry of the \thop ~system.

\section{Discussion}

Our photometric multisite campaign on the \bcep star \thop ~resulted in
the detection of seven independent pulsation modes. Our colour photometry
that was intended for mode typing only resulted in two firm
identifications, but we believe that this is mostly due to the small
photometric amplitudes of all but one of the pulsation modes. However, 
also the strongest mode could not be unambiguously identified from our 
data; we had to invoke literature results.

Mode identification from colour photometry of \bcep stars primarily rests
on the amplitudes determined in the blue and ultraviolet ($\lambda <
4200$~\AA). We have used a subset of the Str\"omgren filter system for our
measurements as a compromise between wide availability and mode
identification potential, with the drawback that the identifications
critically depend on the results in the $u$ filter, which may be hard to
verify. Consequently, any systematic error in the $u$ measurements can
heavily compromise the mode identifications.

The Walraven or the Geneva photometric systems would provide a solution to
this dilemma, but they are unsuitable for multisite work since few
observatories are equipped for their use. However, the pulsation spectrum 
of \thopa ~is reasonably simple and the range of excited frequencies is 
less than 1 \cd, so that an extensive single-site study of this star in 
one of these photometric systems should be sufficient to check the mode 
identifications we finally arrived at by adding further constraints to the 
colour amplitude analysis.

We believe that our suggestion that $f_1, f_5$ and $f_2$ are part of a
rotationally split $\ell=2$ quintuplet can also be checked by theoretical
model calculations. Since the asymmetry of the hypothesised multiplet due
to the second-order effects of rotation has been measured to good relative
accuracy, and since the rotation rate of the star is constrained by
the $f_4, f_6, f_7$ triplet splitting, pulsational models should be able
to reproduce the observed asymmetries if our identification of $f_1, f_5$
and $f_2$ is correct.

To perform a detailed asteroseismic study, one more ambiguity must then
still be overcome: if $f_1, f_5$ and $f_2$ are $\ell=2$ quintuplet
members, their $m$ values must be determined. From photometry alone, we
cannot distinguish if they correspond to $m = (-2, 0, 1)$ or $m = (-1, 1,
2)$. However, the analysis of archival spectroscopy by Briquet et al.
(2005) solves this ambiguity.

It is interesting to note that we found evidence that \thopa ~is seen
close to equator on (a result corroborated by Briquet et al. 2005). In
such a configuration, the $m=0$ component of $\ell=1$ modes as well as the
$|m|=1$ components of $\ell=2$ modes should suffer heavy geometric
cancellation. However, such modes are apparently observed.

In any case, a seismic investigation of \thopa ~is possible. We note that
we would not expect its outcome to be as fruitful as that for $\nu$~Eri
(Pamyatnykh et al. 2004, Ausseloos et al. 2004) since fewer radial
overtones of modes are excited, but the general applicability of earlier
results could be tested. In addition, the detection of a possible eclipse
of the primary would help to constrain the system parameters even tighter,
which can in turn assist the seismic modelling.

Finally, we would like to point out that the frequency structure of \thopa
~is remarkably similar to that of V836 Cen (Aerts et al. 2004a): a radial
mode close to an incomplete $\ell=2$ multiplet of somewhat lower frequency
and a complete $\ell=1$ triplet of higher frequency, and all modes are
contained in a very narrow frequency interval (which is of extremely
similar size in the co-rotating frame). The only differences are the
somewhat higher pulsation frequencies of \thopa ~and its faster rotation.
We thus speculate that once more pulsation spectra of \bcep stars become
known in detail, important clues on mode excitation can be gathered.

\section*{ACKNOWLEDGMENTS}

This work has been supported by the Austrian Fonds zur F\"orderung der
wissenschaftlichen Forschung under grant R12-N02. GH thanks Maryline
Briquet for sharing her results prior to publication, Conny Aerts for
sending a copy of R. Briers' dissertation, Jagoda
Daszy{\'n}ska-Daszkiewicz for supplying some unpublished information and
for comments on a draft version of this paper, Alosha Pamyatnykh for
supplying theoretical instability strip borders for $Z=0.015$, as well as
Luis Balona and Wojtek Dziembowski and his group for permission to use
their computer codes.

\bsp


\begin{thebibliography}{99}

\bibitem[]{}Abt H. A., Levato H., Grosso M., 2002, ApJ 573, 359

\bibitem[]{}Aerts C., Thoul A., Daszy{\'n}ska J., Scuflaire R., Waelkens
C., Dupret M. A., Niemczura E., Noels A., 2003, Sci 300, 1926

\bibitem[]{}Aerts C., et al., 2004a, A\&A 415, 241

\bibitem[]{}Aerts C., et al., 2004b, MNRAS 347, 463

\bibitem[]{}Ausseloos M., Scuflaire R., Thoul A., Aerts C., 2004, MNRAS 
355, 352

\bibitem[]{}Balona L. A., Evers E. A., 1999, MNRAS 302, 349

\bibitem[]{}Breger M., et al., 1993, A\&A 271, 482

\bibitem[]{}Briers R. C., 1971, PhD thesis, Katholieke Universiteit Leuven

\bibitem[]{}Briquet M., Lefever K., Uytterhoeven K., Aerts C., 2005,
MNRAS, in press

\bibitem[]{}Crawford D. L., 1978, AJ 83, 48

\bibitem[]{}Crawford D. L., Barnes J. V., Golson J. C., 1970, AJ 75, 624

\bibitem[]{}Daszy{\'n}ska-Daszkiewicz J., Dziembowski W. A.,
Pamyatnykh A. A., Goupil M.-J., 2002, A\&A 392, 151

\bibitem[]{}Daszy{\'n}ska-Daszkiewicz J., Dziembowski W. A., Pamyatnykh A.
A., 2003, A\&A 407, 999

\bibitem[]{}De Ridder J., et al., 2004, MNRAS 351, 324

\bibitem[]{}Dupret M.-A., Thoul A., Scuflaire R.,
Daszy{\'n}ska-Daszkiewicz J., Aerts C., Bourge P.-O., Waelkens C., Noels
A., 2004, A\&A 415, 251

\bibitem[]{}ESA, 1997, The {\it Hipparcos} and Tycho catalogues, ESA 
SP-1200

\bibitem[]{}Flower P. J., 1996, ApJ 469, 355

\bibitem[]{}Handler G., et al., 2004, MNRAS 347, 454

\bibitem[]{}Handler G., Shobbrook R. R., Vuthela F. F., Balona L. A.,
Rodler F., Tshenye T., 2003, MNRAS 341, 1005

\bibitem[]{}Henroteau F., 1922, Pub. Dom. Obs. Ottawa 8, 1

\bibitem[]{}Heynderickx D., 1992, A\&AS 96, 207

\bibitem[]{}Heynderickx D., Waelkens C., Smeyers P., 1994, A\&AS 105, 447

\bibitem[]{}K\"unzli M., North P., Kurucz R. L., Nicolet B., 1997, A\&AS
122, 51

\bibitem[]{}McAlister H., Mason B. D., Hartkopf W. I., Shara M. M., 1993, 
AJ 106, 1639

\bibitem[]{}Montgomery M. H., O'Donoghue D., 1999, Delta Scuti Star
Newsletter 13, 28 (University of Vienna)

\bibitem[]{}Napiwotzki R., Sch\"onberner D., Wenske V., 1993, A\&A 268,
653

\bibitem[]{}Niemczura E., Daszy{\'n}ska-Daszkiewicz J., 2005, A\&A 433, 
659

\bibitem[]{}Pamyatnykh A. A., 1999, Acta Astr. 49, 119

\bibitem[]{}Pamyatnykh A. A., Dziembowski W. A., Handler G., Pikall H.,
1998, A\&A 333, 141

\bibitem[]{}Pamyatnykh A. A., Handler G., Dziembowski W. A., 2004, MNRAS
350, 1022

\bibitem[]{}Shatsky N., Tokovinin A., 2002, A\&A 382, 92

\bibitem[]{}Sperl M., 1998, Master's Thesis, University of Vienna

\bibitem[]{}Stankov A., Handler G., 2005, ApJS 158, 193

\bibitem[]{}van Hoof A., 1962, Z. Astrophys. 54, 255

\bibitem[]{}van Hoof A., Blaauw A., 1958, ApJ 128, 273

\end{thebibliography}
\end{document}